\documentclass[epj, nopacs, fleqn]{svjour}
\usepackage{graphics}
\usepackage{amsmath}
\usepackage{listings}
\def\slash#1{
\setbox0=\hbox{$#1$}                    
\dimen0=\wd0                            
\setbox1=\hbox{/} \dimen1=\wd1          
\ifdim\dimen0>\dimen1                   
\rlap{\hbox to \dimen0{\hfil/\hfil}}    
#1                                      
\else                                   
\rlap{\hbox to \dimen1{\hfil$#1$\hfil}} 
/                                       
\fi}
\begin{document}
\title{Fast calculation of HELAS amplitudes using graphics processing
unit (GPU)}
\author{K.~Hagiwara\inst{1}
   \and J.~Kanzaki\inst{2}\fnmsep\thanks{e-mail: junichi.kanzaki@kek.jp}
   \and N.~Okamura\inst{2}\fnmsep\thanks{e-mail: naotoshi@post.kek.jp}
   \and D.~Rainwater\inst{3}
   \and T.~Stelzer\inst{4}\fnmsep\thanks{e-mail: tstelzer@uiuc.edu}
   }                     
   %
   %
\institute{KEK Theory Center and Sokendai, Tsukuba 305-0801, Japan
	\and KEK, Tsukuba 305-0801, Japan
	\and Space and Geophysics Laboratory, Applied Research
	Laboratories, University of Texas, Austin, TX 78758, USA
	\and Dept. of Physics, University of Illinois, Urbana, IL, USA
}

\date{Received: date / Revised version: \today}

\abstract{
We use the graphics processing unit (GPU) for fast calculations of helicity
amplitudes of physics processes.
As our first attempt, we compute $u\overline{u}\rightarrow
n\gamma$ ($n\!=\!2$ to 8) processes in $pp$ collisions
at $\sqrt{s} = 14$TeV by transferring the MadGraph generated
HELAS amplitudes (FORTRAN) into newly developed HEGET ({\bf H}ELAS
{\bf E}valuation with {\bf G}PU {\bf E}nhanced {\bf T}echnology) codes
written in CUDA, a C-platform 
developed by NVIDIA for general purpose computing on the GPU.
Compared with the usual CPU programs, we obtain 40-150 times better
performance on the GPU.
} 
		
\titlerunning{Fast calculation of HELAS amplitudes using the GPU}
%

\maketitle

 \section{Introduction}
 \label{sec:intro}
		
  \subsection{Physics motivations}
  
  The field of particle physics is about to enter a new era of experimental
  discovery at the Large Hadron Collider (LHC) which will start collecting data 
  this year.
  Typical new physics signals at the LHC are expected to have many high
  $p_{T}$ jets, $\gamma$'s, $W$'s and $Z$'s, and it is very important to
  estimate the Standard Model (SM) background to all the signals
  reliably.
  Computation of multiple particle production amplitudes is, however,
  often time consuming, and it is desirable to have tools that allow us
  to evaluate cross sections efficiently.
  We examine here the possibility of using a GPU (Graphic Processing Unit)
  to compute helicity amplitudes at many phase space points
  simultaneously, by taking advantage of its parallel processing
  capability. 

  \subsection{GPU hardware and software}

  The GPU is a specialized integrated circuit (IC)
  used in the computer system which outputs complex images onto displays
  very fast.
  This IC is composed of many multi-processors and can process large
  amounts of image data in parallel with high efficiency.
  Because of this high performance GPUs are widely used in the
  scientific applications where a large 
  number of calculations must be performed to process the data.

  Recently NVIDIA~\cite{nvidia} introduced the software
  development system, CUDA\cite{cuda}\footnote{We use CUDA version
  2.1 in this paper.}, which enables one to develop
  programs which can be executed on the GPU using C/C++.
  We use this system to develop a new helicity amplitude calculation
  package which can be used on a GPU.

 \section{Physics process}
 \label{sec:physics}
 
 As our first attempt we compute $u\overline{u}\!\rightarrow\! n\gamma$
 processes in $pp$-collisions, because the amplitudes are particularly
 simple for numerical codes like HELAS~\cite{helas} such that large
 numbers of Feynman diagrams, beyond the capability of automatic diagram 
 generators like MadGraph~\cite{madgraph}, can be computed without 
 difficulty~\cite{ngamma}.  

  \subsection{$n\gamma$ production in $pp$ collisions}
  
\begin{figure}[htb]

   \begin{center}
    \resizebox{0.4\textwidth}{!}{%
    \includegraphics{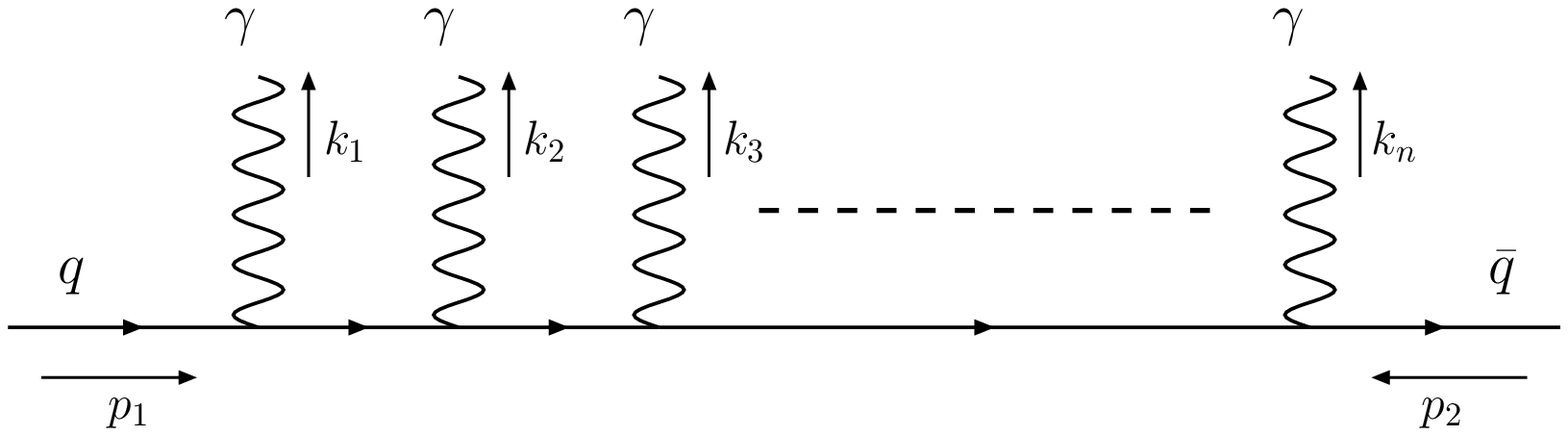}
    }
    \caption{One of $n!$ diagrams for $q\overline{q}\to n\gamma$.}
    \label{fig:uux2na}
   \end{center}
  \end{figure}

  The cross section for producing $n$ high $p_{T}$ photons in $pp$
  collisions can be expressed in the leading order of perturbative QCD as
  \begin{eqnarray}
   {\rm d\sigma}^{n\gamma} &=& \sum_{q}\iint dx_1 dx_2
    \left[
     D_{q/p}\left(x_1,Q\right)
     D_{\bar{q}/p}\left(x_2,Q\right)
	      \right.\nonumber \\
   &&\hspace*{5ex}
    +\left.
      D_{\bar{q}/p}\left(x_1,Q\right)
      D_{q/p}\left(x_2,Q\right)
	 \right]
    {\rm d\hat{\sigma}}\left(\hat{s};y\right)\,,
  \end{eqnarray}
  where the factorization scale $Q$ for the $q$ and $\overline{q}$
  distribution functions is chosen as the $p_\mathrm{T}$ cut-off, and
  the $q\overline{q}\rightarrow n\gamma$'s  process cross section is
  \begin{eqnarray}
   {\rm d\hat{\sigma}}\left(\hat{s}\right)
    =\dfrac{1}{2\hat{s}}
    \dfrac{1}{2^2}
    \dfrac{1}{3^2}\,
    3
    \!\!\!\!
    \sum_{\substack{{\lambda_1,\cdots,\lambda_n}\\{\sigma_1,\sigma_2}}}
    \!\!\!\!
    \left|
     {\cal{M}}^{\lambda_1,\cdots,\lambda_n}_{\sigma_1,\sigma_2}
    \right|^2
    \dfrac{1}{n!}
    ~{\rm d\Phi_n}\,.
  \end{eqnarray}
  Here
  \begin{eqnarray}
   \hat{s}=s\, x_1 x_2\,,
  \end{eqnarray}
  and $\sigma_{i}$($i\!=\!1,2$) denotes the helicity of the initial $u$ and
  $\overline{u}$, $\lambda_{i}$($i\!=\!1$ to n) is that of the final
  photons.
  The $n$-body phase space is
  \begin{eqnarray}
   {\rm d\Phi_n}
    &=&
    \left(2\pi\right)^4
    \delta^4\left(p_1+p_2-\sum_{i=1}^{n}k_i\right)
    \prod_{i=1}^{n}
    \dfrac{d^3k_i}{(2\pi)^{3}\,2\omega_i}\, ,
  \end{eqnarray}
  where $\omega_{i} = |\vec{k}_{i}|$ for massless photons and
  $1/n!$ is the statistical factor for $n$ identical photons.
  The helicity amplitude for the process
  \begin{equation}
  q(p_{1},\sigma_{1}) + \overline{q}(p_{2},\sigma_{2})
  \rightarrow \gamma(k_{1},\lambda_{1}) + \gamma(k_{2},\lambda_{2}) + 
  \cdots + \gamma(k_{n},\lambda_{n})
  \end{equation}
  is particularly simple:
  \begin{eqnarray}
    {\cal{M}}^{\lambda_1,\cdots,\lambda_n}_{\sigma_1,\sigma_2}
   =
   M^{\lambda_1,\cdots,\lambda_n}_{\sigma_1,\sigma_2}
    +(n!-1) \mbox{ permutations}\, ,
    \label{eq:permutation}
  \end{eqnarray}
  where the full amplitude is obtained by $(n!-1)$ permutation of one
  amplitude, denoted e.g. by the Feynman diagram at Fig.~\ref{fig:uux2na}:
  \begin{eqnarray}
   M^{\lambda_1,\cdots,\lambda_n}_{\sigma_1,\sigma_2}&=&
    \bar{v}(p_2,\sigma_2)
    \left[
     \gamma^{\mu_n}
     \dfrac{\slash{p}_1-\slash{K}_{n-1}}{t_{n-1}}
     \gamma^{\mu_{n-1}}
     \cdots
       \right.
    \nonumber\\
   &&\hspace*{3ex}
    \left.
     \cdots
     \gamma^{\mu_3}
     \dfrac{\slash{p}_1-\slash{K}_2}{t_2}
     \gamma^{\mu_2}
     \dfrac{\slash{p}_1-\slash{K}_1}{t_1}
     \gamma^{\mu_1}
	\right] 
    u(p_1,\sigma_1)
    \nonumber\\
   &&\hspace*{3ex}
    \epsilon^{\ast}_{\mu_1}(k_1,\lambda_1)
    \cdots
    \epsilon^{\ast}_{\mu_n}(k_n,\lambda_n)\,(eQ_{q})^{n}\, ,
  \end{eqnarray}
  with
  \begin{equation}
   K_{l} = \sum_{i=1}^{l} k_{i}, \quad 
   t_{l} = \left(p_1-K_l\right)^2\,.
  \end{equation}
  Here $Q_{q}$ is the electric charge of the quark in units of $e$, the
  proton charge.

  The standard version of MadGraph/MadEvent~\cite{madgraph,madevent,newmad}
  generates HELAS amplitudes and computes the cross section
  for the $n\gamma$ production process up to $n\!=\!6$ without
  difficulty.  
  The HELAS amplitude for $n\!=\!7$ with $7\,!$ diagrams can be generated
  by MadGraph after enlarging the maximum size of several variables.
  We are unable to make MadGraph generate $8\,!\approx 4\times 10^{4}$
  diagrams for $n\!=\!8$ photons, so we evaluate them by coding the
  amplitude of eq.~(\ref{eq:permutation}) and by summing over $(8!-\!1)$ 
  permutations of the 8 photon momenta in a numerical program.

  \subsection{Final state cuts}
  \label{subsec:cuts}

  In order to simulate realistic LHC experiments, we introduce final
  state cuts for all the observed photons as follows:
  \begin{equation}
   {\renewcommand\arraystretch{1.2}
   \begin{array}{rcl}
   p_{\mathrm{T}i} &>& p_{\mathrm{T}}^{\scriptstyle\mathrm{cut}} =
    20\,\mathrm{GeV} , \\
   |\eta_{i}| &<& \eta^{\scriptstyle\mathrm{cut}} = 2.5, \\
   \Delta R_{ij} &>& \Delta R^{\scriptstyle\mathrm{cut}} = 0.4, 
   \end{array}
   }
   \label{eq:cuts}
  \end{equation}
  where $\eta_{i}$ is the rapidity of the $i$-th photon and
  \begin{equation} 
   \Delta R_{ij}  \equiv
    \sqrt{(\Delta\eta_{ij})^{2}+(\Delta\phi_{ij})^{2}}
  \end{equation}
  measures rapidity and  azimuthal angle separation between two photons.

  For the sake of simplicity and for the purpose of the present paper,
  we consider only one subprocess, $u\overline{u}\rightarrow
  n\gamma'$s, for the $n$-photon production processes at the LHC, by
  neglecting contributions from $d\overline{d}$, $s\overline{s}$,
  $c\overline{c}$ and $b\overline{b}$ collisions.
  
 \section{Computation on the GPU}
 \label{sec:computation}

 On the GPU, a number of programs can be executed with its
 multi-processors in  parallel.
 These programs which run concurrently must be the same for all
 multi-processors with different input data for each program.
 In order to utilize the high performance of the GPU for computing the
 $n$-photon production processes, we assign a program which processes one
 event on each processor.

 If the program is given a set of random numbers, it generates a phase
 space point and computes a Jacobian and a squared amplitude for the
 phase space point independently from the other programs.
 By keeping the independence of programs among processors their
 structure can become simple, and we can take full advantage of
 the highly parallel computation on the GPU. 

 In this section we describe our software developed for the computation
 of the $n$-photon productions in $pp$ collisions on the GPU and its
 computation environment.
  \begin{table*}[tbh]
   \begin{center}
    \caption{Parameters of GPUs}
    \label{tab:gpu}       
    \begin{tabular}{|c|c|c|c|} \hline
      & GeForce  & GeForce  & GeForce \\ 
      &  GTX280  &  9800GTX &  8800M GTS \\ \hline\hline
     Number of & 30  & 16 & 8 \\
     multiprocessor & & & \\ \hline
     Number of core & 240  & 128 & 64 \\ \hline
     Total amount of  &	 1000  & 500 & 500 \\
     global memory [MB] & & & \\ \hline
     Total amount of  &	 64 & 64 & 64  \\
     constant memory [kB] & & &\\ \hline
     Total amount of shared & 16  & 16 & 16 \\
     memory per block [kB] & & & \\ \hline
     Total number of registers  & 16 & 8 & 8  \\
     available per block [kB] & & & \\ \hline
     Clock rate [GHz] &	1.30  & 1.67 & 0.40 \\
     \hline
    \end{tabular}
   \end{center}
  \end{table*}
 
  \subsection{Structure of the program}
  
  For $n$-photon production processes our program generates momenta and
  helicities of photons in the final state, those of $u$ and
  $\overline{u}$ in the initial state, and computes the total cross
  section of the physics process in the following order: 
  \begin{enumerate}
   \item initialization of the program,
   \item random number generation on the CPU,
   \item transfer random numbers to the GPU,
   \item generate helicities and momenta of $u$, $\overline{u}$ and 
	 $n\gamma$'s using random numbers, and compute
	 squared amplitudes on the GPU, 
   \item transfer the momenta and helicities of external particles,
	 as well as computed squared amplitudes to the CPU, and
   \item sum up all values to obtain the total cross section and
	 distributions on the CPU.
  \end{enumerate}
  Program steps between the generation of random numbers~(2) and the
  summation of computed cross sections~(6) are repeated until we obtain
  enough statistics for the cross section and distributions.

  For the computation of physics quantities for the $n$-photon
  production  processes, we prepare two types of amplitude programs. 
  They are:
  \begin{itemize}
   \item a program converted from the FORTRAN program obtained with the
	 MadGraph~\cite{madgraph}, and
   \item a handwritten CUDA program using permutations of final state
	 photons, via eq.~(\ref{eq:permutation}).
  \end{itemize}
  Both programs are written using the HEGET  ({\bf H}ELAS {\bf
  E}valuation with {\bf G}PU {\bf E}nhanced {\bf T}echnology) functions,
  newly developed codes, written in CUDA~\cite{cuda}, to compute
  helicity amplitudes a la  HELAS~\cite{helas}.

  Sample codes for the case of the 3-photon
  production process are shown in \ref{app:sample} as
  List~\ref{list:uux3amg} 
  and List~\ref{list:uux3aperm} for the converted MadGraph amplitude and
  the permutation amplitude, respectively.
  They compute a sum of amplitudes for all diagrams, \texttt{ampsum}, 
  a complex number, from the
  given momenta, \texttt{\{p1, $\ldots$, p5\}} and helicities,
  \texttt{\{nh1, $\ldots$, nh5\}} of the external particles.
  In the code for the permutation amplitude, List~\ref{list:uux3aperm},
  a function, \texttt{iPNext}, generates a set of integers for the
  next permutation in the sequence. 
  For all computation of amplitudes, new data type ``\texttt{cmplx}'',
  defined in List~\ref{list:cmplx}, for complex numbers on GPU is
  used. 

  \subsection{Execution of a CUDA function on GPU}

  Once a set of random numbers is generated on the CPU,
  computations from the generation of phase space points to the
  calculation of squared amplitudes are done by calling a CUDA function
  which is executed on the GPU. 
  A function, which is called from a CPU program and is executed on a
  GPU, is called a \textit{kernel}.
  In our program, a single call of a kernel function computes the
  scattering amplitudes at multiple phase-space points in parallel on a
  GPU.  

  A unit program, called a \textit{thread}, which is executed on a single
  processor, calculates the amplitudes for one event.
  In the CUDA programming model a set of threads forms a \textit{thread
  block}.
  Threads within one thread block can share data through the
  \textit{shared memory} and cooperate among themselves.
  In the current architecture of NVIDIA's GPUs a thread block can
  contain up to 512 threads.
  The size of a thread block can be changed within this limit when the
  program is executed on a GPU, and we optimize it to obtain the
  best performance of our program.
    
  With a single call of a \textit{kernel}, multiple thread blocks are
  executed in parallel on a GPU.
  A set of thread blocks, which is executed with a single kernel call,
  is called a \textit{grid}.
  Even within a grid, threads in different thread blocks cannot share
  data through the shared memory.
  \begin{table}[hbt]
   \begin{center}
    \caption{Configuration of the host PC}
    \label{tab:host-pc}       
    \begin{tabular}{|c|c|c|} \hline
     & Linux PC & iMac \\ \hline \hline
     CPU & Core2Duo 3GHz & Core2Duo 3.06GHz \\ \hline
     L2 Cache & 6MB & 6MB \\ \hline
     Memory & 4GB & 2GB \\ \hline
     Bus Speed & 1.333GHz & 1.07GHz \\ \hline \hline
     OS & Fedora 8 (64 bit) & MacOS X 10.5.5 \\ 
     \hline \hline 
     GPU & GTX280 \& & 8800M GTS \\
         & 9800GTX & \\ \hline
    \end{tabular}
   \end{center}
  \end{table}

  \subsection{Host PC environment}

  We tested our program on three different GPUs by NVIDIA: the GeForce
  GTX280, 9800GTX and 8800M GTS.
  The parameters for these three GPUs are given in Table~\ref{tab:gpu}.
  The Ge\-Force 9800GTX with 128 processors was introduced in 
  April/2008 by NVIDIA as a high-end single GPU graphic card for the
  high  performance output of complex images like 3D to the PC display.
  The GeForce GTX280 with a new processor architecture, introduced in
  June/2008, has 240 processors, 1GByte global memory and 16k registers,
  which are about twice those of 9800GTX, with a 20\% slower clock rate.
  A GeForce 8800M GTS is installed on an Apple iMac, and  has 64
  processors and the clock rate of 0.4GHz, but still has the same memory
  and register size as the 9800\-GTX.
  \begin{table*}[tb]
   \begin{center}
    \caption{HEGET functions for external lines}
    \label{tab:wavefunc}       
    \begin{tabular}{|c|c|c|} \hline
     External Line & HEGET Function & HELAS Subroutine \\ \hline \hline
     Flowing-In Fermion
     & {\ttfamily ixxxx0}, {\ttfamily ixxxx1}, {\ttfamily ixxxx2}
     & \texttt{IXXXXX}\\
     Flowing-Out Fermion
     & {\ttfamily oxxxx0}, {\ttfamily oxxxx1}, {\ttfamily oxxxx2}
     & \texttt{OXXXXX} \\
     \hline
     Vector Boson
     & {\ttfamily vxxxx0}, {\ttfamily vxxxx1}, {\ttfamily vxxxx2}
     & \texttt{VXXXXX} \\
     \hline
    \end{tabular}
   \end{center}
   \begin{center}
    \caption{List of the HEGET vertex functions}
    \label{tab:vertices}       
    \begin{tabular}{|c|c|c|c|c|} \hline
     Vertex & Inputs & Output & HEGET Function & HELAS Subroutine
     \\ \hline \hline
     FFV & FFV & Amplitude & {\ttfamily iovxxx} & \texttt{IOVXXX} \\
         & FF  & V         & {\ttfamily jioxx0} & \texttt{JIOXXX} \\
         & FV  & F         & {\ttfamily fvixx0}, {\ttfamily fvoxx0} 
         &  \texttt{FVIXXX}, \texttt{FVOXXX}\\
     \hline
    \end{tabular}
   \end{center}
  \end{table*}
  
  Parameters for the two host PCs are summarized in
  Table~\ref{tab:host-pc}.
  They have comparable capability, although the Linux PC has twice larger
  memory size and 25\% faster bus speed.
  It should be noted here that although the host PC's both have
  two CPU's our programs do not run parallel on them.

 \section{HEGET functions}
 \label{sec:heget}

 Based on the FORTRAN version of the HELAS library\cite{helas}
 we developed a set of CUDA functions which can be used on a GPU for
 performing helicity amplitude calculations. 
  These functions are directly converted from the HELAS subroutines into
  C code.
  We kept the order of the arguments of the HELAS subroutines but
  parameters for masses and widths were removed from the
  argument list.

  The HEGET functions are organized to maximize the performance on the
  GPU. 
  Since conditional branches with\-in programs using ``if''
  statements reduces 
  the total efficiency of parallel computing, we eliminated them as much
  as possible in the HEGET functions.
  Accordingly, we prepared separate HEGET codes for the computation of
  massive and massless wave functions. 
  In this paper we only use the HEGET functions for massless particles. 

  By the same token, we introduce three types of functions
  for external wave functions; 
  ``\texttt{1}'' for particles moving in the $+\!z$ direction,
  ``\texttt{2}'' for particles moving in the $-\!z$ direction and
  ``\texttt{0}'' for particles with momentum, $\vec{p}$, along a generic
  direction. 
  These numbers are appended to the name of the corresponding
  functions.
 
  All functions relevant to this paper are listed in
  Table~\ref{tab:wavefunc} and Table~\ref{tab:vertices}, which include
  all functions for massless fermions and massless vector bosons
  (photons), respectively.
  The naming scheme for HEGET functions follow that of HELAS
  subroutines: the HEGET (HELAS) function names start with \texttt{i(I)}
  and \texttt{o(O)} for flow-in and flow-out fermion wave functions,
  respectively, \texttt{v(V)} for vector boson wave functions,
  \texttt{f(F)} for off-shell fermions and \texttt{j(J)} for off-shell
  vector bosons.
  The correspondence between the HEGET functions and the HELAS codes are
  shown explicitly in Tables~\ref{tab:wavefunc} and \ref{tab:vertices}.

  All of the HEGET functions that are used in this report are described
  in \ref{app:heget}.
  
  The massless fermion wave functions with flowing-IN fermion
  number, \texttt{ixxxx\textit{k}} $(k\!=\!0,1,2)$, and those with
  flowing-OUT fermion 
  number, \texttt{oxxxx\textit{k}} $(k\!=\!0,1,2)$, are listed in
  \ref{app:heget-fermion-wave-func}. 
  It is worth noting here that the first 4 components of
  the output complex array \texttt{fi[6]} and \texttt{fo[6]} of these functions
  are 4-spinors
  \begin{subequations}
   \begin{eqnarray}
    \texttt{|fi>} & = & u(\texttt{p},\texttt{nHEL}/2)\mbox{ for
     }\texttt{nSF} = +1 \\ 
    & =  & v(\texttt{p},\texttt{nHEL}/2)\mbox{ for }\texttt{nSF} = -1  
   \end{eqnarray}
  \end{subequations}
  \begin{subequations}
   \begin{eqnarray}
    \texttt{<fo|} & = & \overline{u}(\texttt{p},\texttt{nHEL}/2)\mbox{
     for }\texttt{nSF} = +1 \\ 
    & = &\overline{v}(\texttt{p},\texttt{nHEL}/2)\mbox{ for
     }\texttt{nSF} = -1   
   \end{eqnarray}
  \end{subequations}
  just as in the corresponding HELAS subroutines~\cite{helas}.
  Although either the first two or the latter two components
  vanish for massless fermions, we keep the above generic
  4-spinor form because the efficiency gain achievable by
  avoiding multiplication of zero's is significant only
  for amplitudes with a single massless fermion line with
  many vector boson emissions and it never exceeds a factor
  of two.
  
  The massless vector boson wave functions, \texttt{vxxxx\textit{k}}
  $(k\!=\!0,1,2)$ are listed in \ref{app:heget-vector-wave-func}.
  Here again the first 4 components
  of the output complex array \texttt{vc[6]} give the wave function
  \begin{subequations}
   \begin{eqnarray}
    \texttt{(vc)} & = & \epsilon^\mu(\texttt{p},\texttt{nHEL})^{*}\mbox{ for
     }\texttt{nSV} = +1  \\ 
    & = & \epsilon^\mu(\texttt{p},\texttt{nHEL})\mbox{ for }\texttt{nSV}
     = -1  
   \end{eqnarray}
  \end{subequations}
  in the light-cone gauge
  \begin{equation}
   \epsilon^{\mu} n_{\mu} = 0
  \end{equation}
  with the light-like vector
  \begin{equation}
   n^{\mu} = p_{\mu} = (p^{0}, -p^{1}, -p^{2}, -p^{3})
    \label{eq:light-cone-vect}
  \end{equation}
  just like the HELAS subroutines~\cite{helas} for massless vector
  bosons.
  The use of the light-cone vector~(\ref{eq:light-cone-vect}) allows us
  to express the wave functions solely in terms of the
  four-momenta of the massless vector bosons.

  In \ref{app:heget-ffv-vertex}, we list HEGET functions for
  the \texttt{FFV} vertex
  \newpage
  \begin{equation}
   \mathcal{L}_{\mathrm{F}_{1}\! \mathrm{F}_{2}\! \mathrm{V}}
    \!= \overline{\psi}_{\mathrm{F}_{1}} \gamma^\mu
    \left[ \texttt{gal[0]} \frac{1\!-\!\gamma_{5}}{2}
    \!+\! \texttt{gal[1]} \frac{1\!+\!\gamma_{5}}{2} \right]
    \psi_{\mathrm{F}_{2}} V_{\mu}^{*}
  \end{equation}
  following the HELAS convention~\cite{helas}.
  The amplitude function \texttt{iovxxx}, the off-shell fermion wave
  functions \texttt{fvixx0} and \texttt{fvoxx0}, and the off-shell
  vector wave function \texttt{jioxx0} are listed in \ref{app:heget-ffv-iov},
  B.3.2, B.3.3, and B.3.4, respectively.
  The correspondence between the HEGET functions and
  the original HELAS subroutines are given in Table~\ref{tab:vertices}.

  The complex numbers are defined in each HEGET functions by
  including the header file \texttt{complx.h}, given
  in~\ref{app:heget-complex}, 
  which has been introduced to handle complex numbers on
  CUDA(GPU). 

  
  


  \section{Validation of the HEGET functions}
  \label{sec:validation}

  We have validated all the HEGET functions by comparing the helicity
  amplitudes of each process for many phase space points and for all
  helicity combinations between those computed on GPU with the HEGET
  functions and those computed on CPU with the FORTRAN version of
  HELAS subroutines.  
  For the phase space generation, we use
  MadGraph/MadEvent~\cite{madgraph} and an independent FORTRAN program which
  calculates total cross section and kinematical distributions with
  the Monte Carlo integration program BASES~\cite{bases} as references.
   
   For comparison, we use the same physics parameters as
   MadGraph/MadEvent for all programs. 
   As the parton distribution function, we use CTEQ6L1~\cite{cteq} and 
   set the factorization scale to be the $p_{\mathrm{T}}$ cut of values, 
   $Q = p_{\mathrm{T}}^{\scriptstyle\mathrm{cut}} = 20\,\textrm{GeV}$;
   see eq.~(\ref{eq:cuts}).
   
  \subsection{Total cross sections}
   
   For the calculation of the $n$-photon production cross sections,
   the same final state cuts for all the observed
   photons are applied; see (eq.~(\ref{eq:cuts}) in
   Sec.~\ref{subsec:cuts}). 
   Results for the computation of the total cross sections are listed in
   Table~\ref{tab:sigtot}.
   We find the results obtained by the HEGET
   functions agree with those from the other programs within the statistics
   of generated number of events.
 
   Up to $n\!=\!5$, MadGraph/MadEvent gives the total cross sections and
   distributions, and we find agreements among all the programs.
   As for $n\!=\!6$ and 7, the MadGraph generated HELAS FORTRAN codes can be
   integrated by using BASES~\cite{bases}, and the results agree well
   with those of HEGET codes.
   For $n\!=\!8$, the FORTRAN program that performs $(8!-1)$
   permutations of eq.~(\ref{eq:permutation}) was used to compute
   the matrix elements, which were integrated by BASES.

   \begin{table*}[htb]
    \begin{center}
     \caption{Total cross sections [fb] for $u\overline{u}\to n\gamma$
     at the LHC.}  
     \label{tab:sigtot}       
     \begin{tabular}{ccccl} \hline
      Number of photons & HEGET & Bases & MadGraph/MadEvent &  \\ \hline
     2 & $1.0822\pm 0.0056$
       & $1.08265 \pm 0.00031$ 
       & $1.0811 \pm 0.0019$
       & $\times 10^{4}$ \\
     3 & $6.7776 \pm 0.0082$
       & $6.7849 \pm 0.0051$
       & $6.775 \pm 0.031$ 
       & $\times 10^{0}$ \\
     4 & $1.2400 \pm 0.0030$
       & $1.2280 \pm 0.0029$
       & $1.2372 \pm 0.0041$
       & $\times 10^{-2}$ \\
     5 & $2.598 \pm 0.011$
       & $2.596 \pm 0.010$
       & $2.572 \pm 0.053$
       & $\times 10^{-5}$ \\
     6 & $5.799 \pm 0.017$
       & $5.792 \pm 0.014$
       & ---
       & $\times 10^{-8}$ \\
     7 & $1.264 \pm 0.008$
       & $1.261 \pm 0.004$
       & ---
       & $\times 10^{-10}$ \\
     8 & $2.77 \pm 0.05$ 
       & $2.4 \pm 0.3$
       & ---
       & $\times 10^{-13}$ \\
     \noalign{\smallskip}\hline
     \end{tabular}
    \end{center}
    \end{table*}
 
  \subsection{Kinematical distributions}
  
\begin{figure}[htb]
   \begin{center}
    \resizebox{0.45\textwidth}{!}{%
    \includegraphics{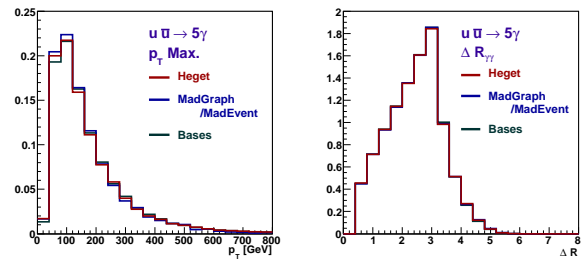}
    }
    \caption{Distributions of maximum transverse momentum and $\Delta R$ of
    final state photons in $u\overline{u}\rightarrow 5\mbox{-photons}$}
    \label{fig:dist}       
   \end{center}
  \end{figure}

  We also compare several kinematical distributions of the generated
  events. 
  As an example, in Fig.~\ref{fig:dist} distributions of the maximum
  transverse momentum and $\Delta R$ between final state photons for the 
  $u\overline{u}\rightarrow 5\mbox{-photons}$ process are shown for the
  three programs.
  We find excellent agreement.
  \break
  Similar comparisons have been performed for all the cases listed in
  Table~\ref{tab:sigtot}, and we find agreement within statistical errors.

  \section{Performance comparison}
  \label{sec:performance}

  \subsection{Register allocation and size of the thread block}
  Event process time is measured by the standard C library function
  on the CPU.
  In order to include the data transfer time between CPU and GPU
  as an integral part of the event process time by GPU, we measure
  the interval between the time when CPU starts transferring random
  numbers to the GPU and the time when the last result from GPU
  is received by the CPU.
  The process time for one event is obtained by dividing the measured time
  by the total number of events.
  We also prepared a CPU program of the same structure and measured the
  event process time for the equivalent part of the program.
  The total number of events used for the measurement depends on the
  number of photons in the final state and the specifications of the GPU.
  For the GTX280, we generate 1M events for a single kernel call and repeated
  100 calls of the kernel for processes with $n_{\gamma}\!\leq\! 5$.
  In short, we generated 100M events for a single execution of the
  program. 

  When we compile CUDA programs, the maximum number of
  registers allocated to a thread can be specified.
  However, the total number of registers available per thread
  block is limited.
  If we allocate more registers to a thread, then the maximum number of
  threads in a thread block, the block size, must become smaller.
  We find the performance of the program depends on these two
  parameters, the number of registers in a thread and the number of
  threads in a block.  Hence, we made a systematic study of the
  performance of these parameters.
 
  The basic unit of the GPU processor is a multiprocessor called
  the Streaming Multiprocessor (SM).
  Threads which belong to one thread block are concurrently executed on
  one SM. 
  One SM consists of eight Scalar Processors (SP) and four
  threads can be executed on one SP at the same time.
  One multiprocessor can therefore process a group of 32 threads, called 
  a \textit{warp}, in parallel. 

  The number of threads in a thread block of a kernel execution can be
  set arbitrarily up to a maximum number limited by the number of
  registers per thread, but the performance is found to be better when
  it is a multiple of 
  the warp size, \textit{i.e.} 32 threads.
  This is because otherwise there are idle SPs during the kernel
  execution. 

\begin{figure}[htb]
   \begin{center}
    \resizebox{0.4\textwidth}{!}{%
    \includegraphics{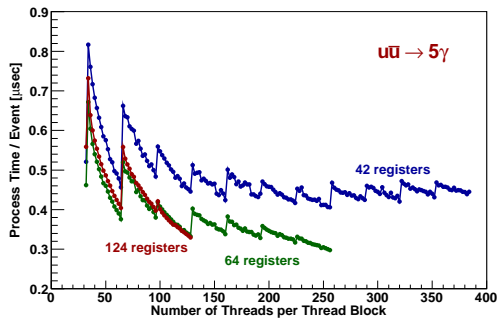}
    }
    \caption{Dependence of event process time on block size (the number
    of threads per thread block) and the number of registers per
    thread on a GTX280.} 
    \label{fig:block}       
   \end{center}
  \end{figure}

  Fig.~\ref{fig:block} shows the dependence of the event process time
  for the 5-photon production process, $u\overline{u}\rightarrow
  5\gamma$, on  the size of thread blocks and the number of allocated
  registers per thread on GTX280. 
  GTX280 has 16k registers, which are split into threads in a thread
  block. 
  Three cases of numbers of registers per thread, 42, 64 and 124, are
  plotted in Fig.~\ref{fig:block}, which correspond to the maximum
  available numbers of threads in a multiple of 32 threads, 384, 256 and
  128, respectively. 

  The dependence on the blocks size shows a clear periodicity of a
  multiple of 32 threads.
  The best performance for the $5\gamma$ production process was
  observed for a combination of 64 registers per thread  
  and 256 threads in  a thread block.
  
  \subsection{Comparison of the event process time}

\begin{figure}[htb]
   \begin{center}
    \resizebox{0.4\textwidth}{!}{%
    \includegraphics{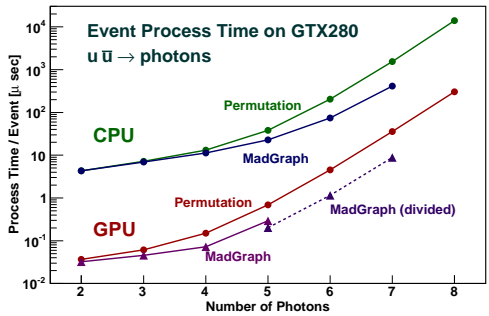}
    }
    \caption{Event process times for the GPU and CPU.}
    \label{fig:time}       
   \end{center}
   \end{figure}

  In Fig.~\ref{fig:time}, the measured process time for one event of
  $n$-photons production 
  processes is shown for the GPU (GTX280) and the CPU (PC Linux with
  Fedora 8). 
  They are plotted versus the number of photons in the final state.

  The upper two lines show the event process time for the computation
  on the CPU. 
  The programs with the MadGraph generated HELAS codes run faster than
  those based on the permutation amplitude for larger number of photons in
  the final state ($n_\gamma\!\geq\! 5$). 
  This is essentially because the MadGraph-generated codes 
  avoid repetition of computing the same off-shell amplitudes.
  A larger fraction of off-shell amplitudes is repeated for larger
  $n_{\gamma}$, and the ratio of the process time for the MadGraph
  generated amplitudes and the permutation amplitudes grows from
  1.7 for $n_{\gamma}\!=\!5$ to 3.8 for $n_{\gamma}\!=\!7$.
  For $n_{\gamma}\!=\! 8$, only the permutation-based program can be
  compiled on the CPU.

  The lower three lines show the event process time on the GPU (GTX280).
  In all cases the program on the GPU computes one event much
  faster than the CPU.
  For the computation on the GPU the HEGET programs obtained by
  converting the HELAS codes generated by the MadGraph
  runs faster than those based on the permutations of a single
  amplitude, when $n_{\gamma}\!=\! 3$ or larger.
  The ratio grows to a factor of 2.3 for $n_{\gamma}\!=\! 5$.

  When the number of photons becomes larger than five, the MadGraph
  amplitude (the HEGET code which is converted from the HELAS code
  generated by MadGraph) becomes very long, and the present CUDA
  compiler cannot process the converted program.
  In order to compile the long HEGET program for large $n$-photon
  processes we divide the program into smaller pieces. 
  Each piece computes a subset of the diagrams.
  The CPU program calls separate kernels for each amplitude
  sequentially and sums up their computed amplitudes.
  This method works for $n\!=\!7$, or $7!\!=\! 5040$ diagrams, whose
  HEGET code can be compiled by dividing it into about 500 kernel
  calls. 
  The 8-photon production process has so many diagrams ($8!\!
  =\!40320$), that we have not been able to compile the HEGET program
  even after dividing it into small pieces.

  For the case of $n_{\gamma}\!\!=\!5$ we examine both programs, one with a 
  single amplitude and the other with divided amplitudes. 
  We find that the multiple kernel program with divided amplitudes runs
  faster than the single kernel program for the whole amplitude by
  about 30\%. 
  This is probably because the size of the $5\gamma$ amplitude program is
  near the maximum capability of the present CUDA compiler. 

  \subsection{Comparison of performance of GPU and CPU}
  
\begin{figure}[htb]
   \begin{center}
    \resizebox{0.4\textwidth}{!}{%
    \includegraphics{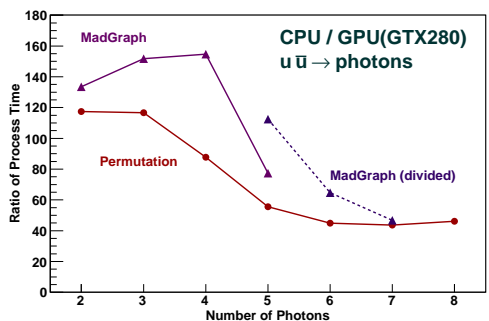}
    }
    \caption{Ratio of event process times (CPU/GPU).}
    \label{fig:gpu-cpu}       
   \end{center}
  \end{figure}

  The ratios of event process times between the CPU and GPU are shown in
  Fig.~\ref{fig:gpu-cpu}.
  The solid curves give the ratios between the permutation amplitudes on
  the CPU (the upper CPU points in Fig.~\ref{fig:time}) and those on the
  GPU (the upper GPU points in Fig.~\ref{fig:time}).
  The performance ratio is about 120 for $n_{\gamma}\!=\!2$ and 3,
  and it goes down to about 45 for $n_{\gamma}\!=\!6$, 7 and 8.
  A higher-performance ratio is found when we compare the event process
  time of the MadGraph-generated amplitude on the CPU (the lower CPU points in
  Fig.~\ref{fig:time}) and that on the GPU (the lower GPU points in
  Fig.~\ref{fig:time}).
  Ratios above 150 are found for $n_{\gamma}\!=\!3$ and 4,
  going down to about 45 for $n_{\gamma}\!=\!7$.
  For $n_{\gamma}\!=\!6$ and 7, the HEGET programs from the MadGraph-generated
  HELAS codes are too big for the present CUDA compiler, and
  the results are shown for the GPU program with multiple kernel calls
  for one event.
  Nevertheless, the performance ratio is about 60 for $n_{\gamma}\!=\!6$
  and about 45 for $n_{\gamma}\!=\! 7$.
  Even more surprisingly, the performance is better for the divided
  amplitude with multiple kernel calls than the single kernel
  computation for $n_{\gamma}\!=\!5$, with the performance ratio
  exceeding 100.

  To summarize, we find that GPU programs run faster than CPU
  programs by a factor exceeding 100 for $n_{\gamma}\leq\! 5$, while a
  performance gain larger than 40 is achieved for other $n_{\gamma}\!\leq\! 8$.

  \subsection{Difference among GPUs}

\begin{figure}[htb]
   \begin{center}
    \resizebox{0.4\textwidth}{!}{%
    \includegraphics{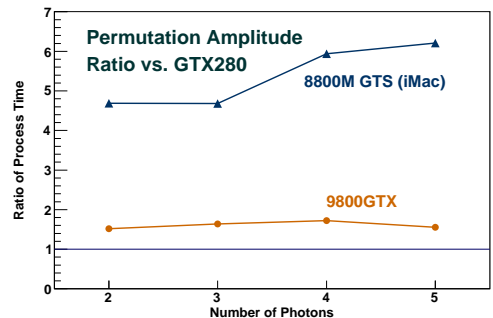}
    }
    \caption{Event process times for GPUs.}
    \label{fig:gpu}       
   \end{center}
  \end{figure}

  We also examine the performance of two other GPUs, the 9800GTX and the
  8800M GTS. 
  The GPU board with 9800GTX is connected to the same host PC as
  GTX280. 
  On the other hand, the 8800M GTS is a built-in graphic card on the
  iMac. 
  We ran the same programs for all GPUs and measures their event process
  times. 
  Fig.~\ref{fig:gpu} shows the process time per event for the
  permutation amplitudes on the 9800GTX and 8800M GTS relative to a GTX280, 
  which was shown in Fig.~\ref{fig:time} as ``GPU permutation''.  
  The ratio of process times among various GPU's is roughly
  proportional to the inverse of the number of multiprocessors: 240 on the
  GTX280, 128 on the 9800GTX, 64 on the 8800M GTS; see Table~\ref{tab:gpu}.  
  It is worth noting that even the small built-in graphic card (8800M
  GTS) on an iMac can compute $n_{\gamma}$ production processes faster than
  CPU by about factors of 25, 15 and 10, respectively, for
  $n_{\gamma}\!=\!3$, 4 and 5.

  \subsection{Double precision calculation}

\begin{figure}[htb]
   \begin{center}
    \resizebox{0.4\textwidth}{!}{%
    \includegraphics{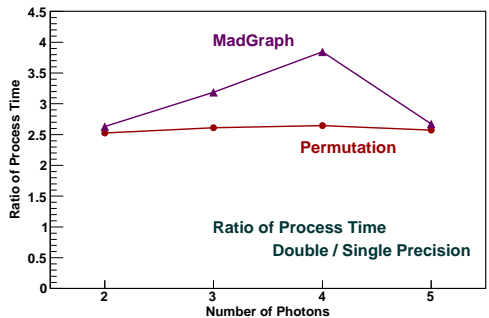}
    }
    \caption{Ratio of event process times (double/single precision) on a GTX280.}
    \label{fig:double}       
   \end{center}
  \end{figure}

  The GTX280 has one double precision processing unit for each
  streaming multiprocessor (SM).
  We prepared a double precision version of the HEGET library and the
  CUDA program which computes the $n_{\gamma}$ cross sections with
  double precision.
  The structure of these programs is the same as the single precision
  version described above.
  Results of the single and double precision computations agree.
  Fig.~\ref{fig:double} shows the ratio of event process times
  between double and single precision computations.
  Because the number of  double precision processing units is
  small compared to the number of single precision processing units, the program
  with the double precision computations runs 3-4 times slower than that
  with single precision.
  In comparison, there is little difference between single and
  double precision computation time on CPU.  

  \subsection{Effect of unrolling loops}
  
  In the permutation amplitude program, a combination of integer numbers
  for successive permutations is generated sequentially 
  within a \textit{while} loop of the C-language: see
  \ref{app:sample-loop} for $n_{\gamma}\!=\!3$ as an example.
  Since programs with loops or branches lower the efficiency of
  parallel computing on the GPU, in general, there is a possibility of
  improving the performance of the permutation program by spreading out,
  or unrolling, a part of the \textit{while} loop in the permutation.

\begin{figure}[htb]
   \begin{center}
    \resizebox{0.4\textwidth}{!}{%
    \includegraphics{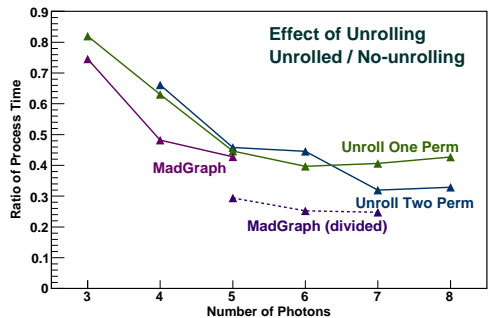}
    }
    \caption{Effect of unrolling \textit{while} loops in ``permutation''
    amplitudes on a GTX280.}
    \label{fig:unroll}       
   \end{center}
  \end{figure}

  We therefore tested the effect of unrolling the \textit{while} loop for 
  the permutation generation. 
  Sample code for the case of the 3-photon
  production process is shown in \break
  \ref{app:sample-unroll} as
  List~\ref{list:uux3aunroll}, 
  for the amplitude program with one permutation unrolled.
  It should be compared with the program List~\ref{list:uux3aperm} in
  \ref{app:sample-loop}, which performs $3!\!=\!6$ permutation within
  one \textit{while} loop.
  Fig.~\ref{fig:unroll} shows the relative performance of the programs
  after unrolling a part of the \textit{while} loop as compared to the
  process time presented in Fig.~\ref{fig:time} as ``GPU permutation''.

  We find that the effect of unrolling is significant, especially for
  large $n_{\gamma}$ processes.
  The improvement of the execution speed becomes a factor of 3 for
  $n_{\gamma}\!=\!7$ and 8.
  Because the event process time does not change significantly by
  unrolling the \textit{while} loop on the CPU, the CPU/GPU ratio
  plotted in Fig.~\ref{fig:gpu-cpu} for `permutation` amplitudes grows
  from $\sim 45$ to $\sim 150$ for $n_{\gamma}\!=\!7$ and 8 after 
  unrolling two permutations.

  Also plotted as a reference in Fig.~\ref{fig:unroll} are the relative
  performances of the computation with the converted
  MadGraph amplitudes as compared to those of permutation amplitudes.
  It is clearly seen from Fig.~\ref{fig:unroll} that by reducing the
  \textit{while} loop the performance of the program with permutation
  improves 
  significantly for $n_{\gamma}\!\geq\! 5$, approaching that of the
  MadGraph based programs.
  The presence of the \textit{while} loop is probably the main cause of
  the poor performance of 
  the permutation amplitudes as compared to the MadGraph generated ones,
  as shown in Fig.~\ref{fig:time}.

 \section{Summary}
 \label{sec:summary}

 We have presented the results of our attempt to compute multi-particle
 production events at hadron colliders on GPUs~\cite{nvidia},
 Graphic Processing Units.
 Our achievements and findings may be summarized as follows.
 \begin{itemize}
  \item HELAS subroutines~\cite{helas} written in FORTRAN were
        converted to HEGET functions written in CUDA~\cite{cuda},
        a C-language developed for GPU computing.
  \item The HELAS amplitude code for $u\overline{u}\rightarrow
	n\gamma$'s ($n_{\gamma}\!\leq\! 7$) generated by
	MadGraph~\cite{madgraph} was converted to a CUDA program which
	calls HEGET functions.
  \item CUDA programs that compute ($n\,!\!-\!1$) permutations of a
	single Feynman amplitude were also made for $n_{\gamma}\!\leq\!8$.
  \item Event process times of the GPU program on a GTX280 are more than 100
	times faster than the CPU program for $n_{\gamma}\!\leq\!5$,
	while the gain is reduced to about 45 for $n_{\gamma}\!=\!7$ and
	8. 
  \item The present CUDA compiler cannot process the \break HEGET programs
	converted from the MadGraph generated HELAS codes for
	$n_{\gamma}\!\geq\!6$. 
  \item For $n_{\gamma}\!=\!5$, 6 and 7, we were successful in
	running the converted MadGraph programs by dividing the whole
	amplitude into smaller pieces and calling them sequentially on
	the CPU.  
	Performance improves by this dividing procedure for
	$n_{\gamma}\!=\!5$, where the whole amplitude can be computed by
	one kernel call.
  \item CUDA programs based on permutations of one HEGET amplitude
	are about a factor of 3 to 4 slower than those based on divided
	MadGraph codes for $n_{\gamma}\!\geq\!5$.
	The main cause of this slowdown was identified as the
	\textit{while} loop in the program that generates permutation of 
	$n_{\gamma}$-integers.
  \item The same programs run on the GTX280, 9800GTX and 8800M GTS, and
	the performance is roughly proportional to the number of
	processors: 240, 128 and 64, respectively. 
  \item Double precision amplitudes on a GTX280 were examined,
	finding a factor of 2.5 to 4 slower performance compared to
	the single precision computation.
 \end{itemize}
 
 \begin{acknowledgement}{\textit{Acknowledgement}.}
  We thank Johan Alwall, Qiang Li and Fabio Maltoni for stimulating
  discussions. 
  This work is supported  by  the Grant-in-Aid for Scientific
  Research from the Japan Society 
  for the Promotion of Science (No. 20340064) and
  the National Science Foundation (No. 0757889).
 \end{acknowledgement}
 
 \appendix
 \def\thesection{Appendix \Alph{section}}
 
\section{Sample codes for amplitude calculations}\label{app:sample}

 \subsection{Three photon production amplitude with the converted MadGraph code}\label{app:sample-mg}

\begin{lstlisting}[caption=uux3a(madgraph).cu,label=list:uux3amg]
        cmplx w01[6], w02[6], w03[6], w04[6], w05[6];

        ixxxx1(p1, nh1, +1, w01);
        oxxxx2(p2, nh2, -1, w02);
        vxxxx0(p3, nh3, +1, w03);
        vxxxx0(p4, nh4, +1, w04);
        vxxxx0(p5, nh5, +1, w05);

        cmplx w06[6], w07[6], w08[6];
        
        cmplx ampsum = mkcmplx(0.0f, 0.0f);
        cmplx amp;
        
        fvoxx0(w02,w03,gau,w06);
        fvoxx0(w06,w04,gau,w07);
        iovxxx(w01,w07,w05,gau,amp);
        ampsum += amp;
        fvixx0(w01,w04,gau,w07);
        fvoxx0(w02,w05,gau,w08);
        iovxxx(w07,w08,w03,gau,amp);
        ampsum += amp;
        fvoxx0(w02,w03,gau,w06);
        fvixx0(w01,w04,gau,w07);
        iovxxx(w07,w06,w05,gau,amp);
        ampsum += amp;
        fvoxx0(w02,w04,gau,w06);
        fvixx0(w01,w05,gau,w07);
        iovxxx(w07,w06,w03,gau,amp);
        ampsum += amp;
        fvixx0(w01,w03,gau,w07);
        fvixx0(w07,w04,gau,w08);
        iovxxx(w08,w02,w05,gau,amp);
        ampsum += amp;
        fvixx0(w01,w03,gau,w07);
        fvoxx0(w02,w04,gau,w06);
        iovxxx(w07,w06,w05,gau,amp);
        ampsum += amp;
\end{lstlisting}

 \subsection{Three photon production amplitude with permutations}\label{app:sample-loop}

\begin{lstlisting}[caption=uux3a(permutation).cu,label=list:uux3aperm]
        cmplx w[5][6];

        ixxxx1(p1, nh1, +1, w[0]);
        oxxxx2(p2, nh2, -1, w[1]);
        vxxxx0(p3, nh3, +1, w[2]);
        vxxxx0(p4, nh4, +1, w[3]);
        vxxxx0(p5, nh5, +1, w[4]);

        cmplx w01[6],w02[6];
        
        cmplx ampsum = mkcmplx(0.0f, 0.0f);
        
        int Flag_Perm=1;
        int a[3];
        a[0] = 2;
        a[1] = 3;
        a[2] = 4;
        
        do {
            cmplx amp;
            fvoxx0(w[1], w[a[0]], gau, w01);
            fvoxx0(w01 , w[a[1]], gau, w02);
            iovxx0(w[0], w02, w[a[2]], gau, amp);
            ampsum += amp;
            iPNext(a, 3, Flag_Perm);
        } while (Flag_Perm>0);
\end{lstlisting}

 \section{HEGET functions}
 \label{app:heget}

  All of the HEGET functions that are used in this report are explained
  in this Appendix.

 \subsection{Fermion wave functions}
 \label{app:heget-fermion-wave-func}

  We have two types of functions to compute external fermi\-ons.
  One is for ``flowing-In'' fermions and the other is for ``flowing-Out''
  fermions.

  For the flowing-In spinor wavefunction of a fermion, we prepare
  three functions, \texttt{ixxxx1} (List \ref{list:ixxxx1}),
  \texttt{ixxxx2} (List \ref{list:ixxxx2}) and
  \texttt{ixxxx0} (List \ref{list:ixxxx0}),
  derived from the HELAS subroutine, \texttt{IXXXXX}; see
  Table~\ref{tab:wavefunc} for correspondence between the HEGET
  functions and the HELAS subroutines~\cite{helas}.
  These three functions compute massless fermion wavefunctions with
  the flowing-In fermion number.
  \texttt{ixxxx1} is for a fermion with momentum along the $+\!z$
  direction, 
  \texttt{ixxxx2} is for a fermion with momentum along the $-\!z$
  direction,  and
  \texttt{ixxxx0} is for a fermion with a generic three-momentum,
  $\vec{p}$. 
  All functions have the same argument list:
  \begin{equation}
     \mbox{\ttfamily ixxxx\textit{k}(float* p, int nHEL, int nSF, cmplx* fi)} 
  \end{equation}
  for $k = \texttt{1}, \texttt{2}$ and $\texttt{0}$.
  The inputs and outputs are:
  \begin{equation}
   \begin{array}{ll}
    \textsc{Inputs:} & \\
     \texttt{float p[4]} & \textrm{4-momentum} \\
     \texttt{int nHEL}   & \textrm{twice fermion helicity (-1 or 1)} \\
     \texttt{int nSF}    & \textrm{+1 for particle, -1 for
      anti-particle} \\ \\
    \textsc{Outputs:} & \\
    \texttt{cmplx fi[6]} & \textrm{fermion wavefunction \texttt{|fi>}} \\
    & u(\texttt{p},\texttt{nHEL}/2) \textrm{ for } \texttt{nSF} = +1 \\
    & v(\texttt{p},\texttt{nHEL}/2) \textrm{ for } \texttt{nSF} = -1 \\
    \label{eq:i} 
   \end{array}
  \end{equation}   

  For the flowing-Out spinor wavefunction of a fermion, we also prepare
  three functions, \texttt{oxxxx1} (List \ref{list:oxxxx1}),
  \texttt{oxxxx2} (List \ref{list:oxxxx2}) and
  \texttt{oxxxx0} (List \ref{list:oxxxx0}),
  derived from the HELAS subroutine, \texttt{OXXXXX}; see
  Table~\ref{tab:wavefunc}. 

  These three functions compute a massless fermion wavefunction with
  a flowing-Out fermion number.
  \texttt{oxxxx1} is for a fermion with momentum along the $+\!z$
  direction,  \texttt{oxxxx2} is for a fermion with momentum along
  $-\!z$ direction,  and
  \texttt{oxxxx0} is for a final state fermion with a generic
  three-momentum, $\vec{p}$. 
  All functions have the same argument list:
  \begin{equation}
   \texttt{oxxxx\textit{k}(float* p, int nHEL, int nSF, cmplx* fo)} \\
  \end{equation}
  for $k = \texttt{1}, \texttt{2}$ and $\texttt{0}$.
  The inputs and outputs are:
  \begin{equation}
   \begin{array}{ll}
    \textsc{Inputs:} & \\
     \texttt{float p[4]} & \textrm{4-momentum} \\
     \texttt{int nHEL}   & \textrm{twice fermion helicity (-1 or 1)} \\
     \texttt{int nSF}    & \textrm{+1 for particle, -1 for
      anti-particle} \\ \\
    \textsc{Outputs:} & \\
    \texttt{cmplx fo[6]} & \textrm{fermion wavefunction
     \texttt{<fo|}} \\
    & \overline{u}(\texttt{p},\texttt{nHEL}/2) \textrm{ for }
     \texttt{nSF} = +1 \\ 
    & \overline{v}(\texttt{p},\texttt{nHEL}/2) \textrm{ for }
     \texttt{nSF} = -1 \\ 
   \end{array}
  \end{equation}   

\begin{lstlisting}[caption=ixxxx1.cu,label=list:ixxxx1]
#include "cmplx.h"

__device__
void ixxxx1(float* p, int nHEL, int nSF, cmplx* fi)
{
    float SQP0P3 = sqrtf(p[0]+p[3])*(float)(nSF);
    int   NH = nHEL*nSF;

    fi[4] = mkcmplx(p[0]*(float)(nSF), 
                    p[3]*(float)(nSF));
    fi[5] = mkcmplx(p[1]*(float)(nSF), 
                    p[2]*(float)(nSF));

    cmplx CHI     = mkcmplx(NH*p[1]*(1.0f/SQP0P3),
                               p[2]*(1.0f/SQP0P3));
    cmplx CZERO   = mkcmplx(0.0f,   0.0f);
    cmplx CSQP0P3 = mkcmplx(SQP0P3, 0.0f);

    fi[0]=(NH== 1)*CZERO   + (NH==-1)*CHI;
    fi[1]=(NH== 1)*CZERO   + (NH==-1)*CSQP0P3;
    fi[2]=(NH== 1)*CSQP0P3 + (NH==-1)*CZERO;
    fi[3]=(NH== 1)*CHI     + (NH==-1)*CZERO;
            
    return;
}
\end{lstlisting}
\begin{lstlisting}[caption=ixxxx2.cu,label=list:ixxxx2]
#include "cmplx.h"

__device__
void ixxxx2(float* p, int nHEL, int nSF, cmplx* fi)
{
    int     NH=nHEL*nSF;

    fi[4] = mkcmplx(p[0]*(float)(nSF),
                    p[3]*(float)(nSF));
    fi[5] = mkcmplx(p[1]*(float)(nSF),
                    p[2]*(float)(nSF));

    cmplx CHI     = mkcmplx( nHEL*sqrtf(2.0f*p[0]),
                             0.0f);
    cmplx CZERO   = mkcmplx(0.0f,0.0f);

    fi[0]=(NH== 1)*CZERO   + (NH==-1)*CHI;
    fi[1]=CZERO;
    fi[2]=CZERO;
    fi[3]=(NH== 1)*CHI     + (NH==-1)*CZERO;
            
    return;
}
\end{lstlisting}
\begin{lstlisting}[caption=ixxxx0.cu,label=list:ixxxx0]
#include "cmplx.h"

__device__
void ixxxx0(float* p, int nHEL, int nSF, cmplx* fi)
{
    float SQP0P3 = sqrtf(p[0]+p[3])*(float)(nSF);
    int   NH = nHEL*nSF;

    fi[4] = mkcmplx(p[0]*(float)(nSF), 
                    p[3]*(float)(nSF));
    fi[5] = mkcmplx(p[1]*(float)(nSF), 
                    p[2]*(float)(nSF));

    cmplx CHI     = mkcmplx( NH*p[1]*(1.0f/SQP0P3),
                             (1.0f/SQP0P3)*p[2]);
    cmplx CZERO   = mkcmplx(0.0f,0.0f);
    cmplx CSQP0P3 = mkcmplx(SQP0P3, 0.0f);

    fi[0]=(NH== 1)*CZERO   + (NH==-1)*CHI;
    fi[1]=(NH== 1)*CZERO   + (NH==-1)*CSQP0P3;
    fi[2]=(NH== 1)*CSQP0P3 + (NH==-1)*CZERO;
    fi[3]=(NH== 1)*CHI     + (NH==-1)*CZERO;

    return;
}
\end{lstlisting}

\begin{lstlisting}[caption=oxxxx1.cu,label=list:oxxxx1]
#include "cmplx.h"

__device__
void oxxxx1(float* p, int nHEL, int nSF, cmplx* fo)
{
    float   SQP0P3=sqrtf(p[0]+p[3])*(float)(nSF);
    int     NH=nHEL*nSF;

    fo[4] = mkcmplx(p[0]*(float)(nSF),
                    p[3]*(float)(nSF));
    fo[5] = mkcmplx(p[1]*(float)(nSF),
                    p[2]*(float)(nSF));
    
    cmplx CHI     = mkcmplx( NH*p[1]*(1.0f/SQP0P3),
                             -p[2]*(1.0f/SQP0P3));
    cmplx CZERO   = mkcmplx(0.0f,   0.0f);
    cmplx CSQP0P3 = mkcmplx(SQP0P3, 0.0f);

    fo[0]=(NH== 1)*CSQP0P3 + (NH==-1)*CZERO;
    fo[1]=(NH== 1)*CHI     + (NH==-1)*CZERO;
    fo[2]=(NH== 1)*CZERO   + (NH==-1)*CHI;
    fo[3]=(NH== 1)*CZERO   + (NH==-1)*CSQP0P3;

    return;
}
\end{lstlisting}
\begin{lstlisting}[caption=oxxxx2.cu,label=list:oxxxx2]
#include "cmplx.h"

__device__
void oxxxx2(float* p, int nHEL, int nSF, cmplx* fo)
{
    int     NH=nHEL*nSF;

    fo[4] = mkcmplx(p[0]*(float)(nSF),
                    p[3]*(float)(nSF));
    fo[5] = mkcmplx(p[1]*(float)(nSF),
                    p[2]*(float)(nSF));
    
    cmplx CHI     = mkcmplx(-nHEL*sqrtf(2.0f*p[0]),
                            0.0f);
    cmplx CZERO   = mkcmplx(0.0f,0.0f);

    fo[0]=CZERO;
    fo[1]=(NH== 1)*CHI     + (NH==-1)*CZERO;
    fo[2]=(NH== 1)*CZERO   + (NH==-1)*CHI;
    fo[3]=CZERO;

    return;
}
\end{lstlisting}
\begin{lstlisting}[caption=oxxxx0.cu,label=list:oxxxx0]
#include "cmplx.h"

__device__
void oxxxx0(float* p, int nHEL, int nSF, cmplx* fo)
{
    float SQP0P3 = sqrtf(p[0]+p[3])*(float)(nSF);
    int   NH = nHEL*nSF;

    fo[4] = mkcmplx(p[0]*(float)(nSF), 
                    p[3]*(float)(nSF));
    fo[5] = mkcmplx(p[1]*(float)(nSF),
                    p[2]*(float)(nSF));
    
    cmplx CHI     = mkcmplx( NH*p[1]*(1.0f/SQP0P3),
                             -p[2]*(1.0f/SQP0P3));
    cmplx CZERO   = mkcmplx(0.0f,   0.0f);
    cmplx CSQP0P3 = mkcmplx(SQP0P3, 0.0f);

    fo[0]=(NH== 1)*CSQP0P3 + (NH==-1)*CZERO;
    fo[1]=(NH== 1)*CHI     + (NH==-1)*CZERO;
    fo[2]=(NH== 1)*CZERO   + (NH==-1)*CHI;
    fo[3]=(NH== 1)*CZERO   + (NH==-1)*CSQP0P3;

    return;
}
\end{lstlisting}
 
 \subsection{Vector boson wave functions}
 \label{app:heget-vector-wave-func}

  For the wavefunction of a massless vector boson, we prepare
  three functions, \texttt{vxxxx1} (List~\ref{list:vxxxx1}),
  \texttt{vxxxx2} (List~\ref{list:vxxxx2}) and
  \texttt{vxxxx0} (List~\ref{list:vxxxx0}),
  derived from the HELAS subroutine, \texttt{VXXXXX}; see
  Table~\ref{tab:wavefunc}. 

  These three functions compute a massless vector boson wavefunction
  from its four-momentum and helicity.
  \texttt{vxxxx1} is for a vector boson with momentum along the $+\!z$
  direction, \texttt{vxxxx2} is for a vector boson with momentum
  along $-\!z$ direction,  and
  \texttt{vxxxx0} is for a vector boson with a generic
  three-momentum, $\vec{p}$. 
  Their argument list is:
  \begin{equation} 
    \texttt{vxxxx\textit{k}(float* p, int nHEL, int nSV, cmplx* vc)} 
  \end{equation}
  for $k = \texttt{1}, \texttt{2}$ and $\texttt{0}$.
  The inputs and outputs are:
  \begin{equation}
   \begin{array}{ll}
    \textsc{Inputs:} & \\
     \texttt{float p[4]} & \textrm{4-momentum} \\
     \texttt{int nHEL}   & \textrm{helicity of massless vector (-1 or 1)} \\
     \texttt{int nSV}    & \textrm{+1 for final, -1 for initial
      vector} \\ \\
    \textsc{Outputs:} & \\
    \texttt{cmplx vc[6]} & \textrm{vector boson wavefunction} \\
    & \epsilon^{\mu}(p, \texttt{nHel})^{*} \textrm{ for }
     \texttt{nSV} = +1 \\ 
    & \epsilon^{\mu}(p, \texttt{nHel}) \textrm { for }
     \texttt{nSV} = -1.
   \end{array}
  \end{equation}   

\begin{lstlisting}[caption=vxxxx1.cu,label=list:vxxxx1]
#include "cmplx.h"

__device__
void vxxxx1(float* p, int nHEL, int nSV, cmplx* vc)
{
    vc[4] = mkcmplx(-p[0], -p[3]);
    vc[5] = mkcmplx(-p[1], -p[2]);

    vc[0] = mkcmplx(0.0f, 0.0f);
    vc[3] = mkcmplx(0.0f, 0.0f);

    vc[1] = mkcmplx(-(float)(nHEL)*rSQRT2,    0.0f);
    vc[2] = mkcmplx(                 0.0f, -rSQRT2);

    return;
}
\end{lstlisting}
\begin{lstlisting}[caption=vxxxx2.cu,label=list:vxxxx2]
#include "cmplx.h"

__device__
void vxxxx2(float* p, int nHEL, int nSV, cmplx* vc)
{
    vc[4] = mkcmplx(-p[0], -p[3]);
    vc[5] = mkcmplx(-p[1], -p[2]);

    vc[0] = mkcmplx(0.0f, 0.0f);
    vc[3] = mkcmplx(0.0f, 0.0f);

    vc[1] = mkcmplx(-(float)(nHEL)*rSQRT2,    0.0f);
    vc[2] = mkcmplx(                 0.0f,  rSQRT2);

    return;
}

\end{lstlisting}
\begin{lstlisting}[caption=vxxxx0.cu,label=list:vxxxx0]
#include "cmplx.h"

__device__
void vxxxx0(float* p, int nHEL, int nSV, cmplx* vc)
{
    vc[4] = mkcmplx(p[0],  p[3])*nSV;
    vc[5] = mkcmplx(p[1],  p[2])*nSV;

    float rpt = rsqrtf(p[1]*p[1] + p[2]*p[2]);

    vc[0] = mkcmplx(0.0f,  0.0f);
    vc[3] = mkcmplx(
        (float)(nHEL)*(1.0f/(rpt*p[0]))*rSQRT2,
        0.0f);

    float pzpt = (p[3]*(1.0f/p[0])*rpt)*rSQRT2
        *(float)(nHEL);
    
    vc[1] = mkcmplx(-p[1]*pzpt,
                    -nSV*p[2] * rpt * rSQRT2);
    vc[2] = mkcmplx(-p[2]*pzpt,
                    +nSV*p[1] * rpt * rSQRT2);

    return;
}
\end{lstlisting}

 \subsection{Fermion-fermion-vector boson vertex}
 \label{app:heget-ffv-vertex}
 
   For the fermion-fermion-vector boson vertex, 
  \begin{equation}
   \mathcal{L}_{\mathrm{F}_{1}\! \mathrm{F}_{2}\! \mathrm{V}}
    \!= \overline{\psi}_{\mathrm{F}_{1}} \gamma^\mu
    \left[ \texttt{gal[0]} \frac{1\!-\!\gamma_{5}}{2}
    \!+\! \texttt{gal[1]} \frac{1\!+\!\gamma_{5}}{2} \right]
    \psi_{\mathrm{F}_{2}} V_{\mu}^{*}
  \end{equation}
  there are four functions,
  \texttt{iovxxx}, \texttt{fvixx0}, \texttt{fvoxx0} and \break\texttt{jioxx0},
  in HEGET.
  They correspond to HELAS subroutines, \texttt{IOVXXX},
  \texttt{FVIXXX}, \texttt{FVOXXX} and \texttt{JIOXXX}, respectively,
  for massless particles; see Table~\ref{tab:vertices} for the
  correspondence between the HEGET functions and the HELAS
  subroutines~\cite{helas}. 

  \subsubsection{\texttt{iovxxx}}
  \label{app:heget-ffv-iov}

  The function \texttt{iovxxx} (List~\ref{list:iovxxx})
  computes the amplitude of 
  the \texttt{FFV} vertex from a flowing-In fermion,
  a flowing-Out fermion and a vector boson wave functions, whether they
  are on-shell or off-shell.
  It has the arguments:
  \begin{equation} 
   \begin{array}{r}
   \texttt{iovxxx(cmplx* fi, cmplx* fo, cmplx* vc, } \\
   \texttt{float* gal, cmplx vertex)} 
    \end{array}
  \end{equation}
  where the inputs and the outputs are:
  \begin{equation}
   \begin{array}{ll}
    \textsc{Inputs:} & \\
     \texttt{cmplx fi[6]} & \textrm{flowing-In fermion wavefunction} \\
     \texttt{cmplx fo[6]} & \textrm{flowing-Out fermion wavefunction} \\
     \texttt{cmplx vc[6]} & \textrm{vector wavefunction} \\
     \texttt{float gal[2]} & \textrm{coupling constants of \texttt{FFV}
      vertex} \\ \\
    \textsc{Outputs:} & \\
    \texttt{cmplx vertex} & \textrm{amplitude of the \texttt{FFV} 
     vertex} \\
    & \qquad\qquad\qquad\qquad\qquad \texttt{<fo|V|fi>}
     \label{eq:iov}
    \end{array}
  \end{equation}
  The two chiral couplings of eq.~(\ref{eq:iov}) are:
  \begin{equation}
   \texttt{gal[0]} = \texttt{gal[1]} = -eQ_{q}
  \end{equation}
  for the $qq\gamma$ vertex in QED, where $Q_{q}$ is the electric charge
  of the quark in units of the proton charge, $e=\sqrt{4\pi\alpha}$.

  \subsubsection{\texttt{fvixx0}}
  \label{app:heget-ffv-fvi}

  The function \texttt{fvixx0} (List.~\ref{list:fvixx0})
  computes the off-shell massless fermion wavefunction
  from a flowing-In external fermi\-on and a vector boson.
  It has the arguments:
  \begin{equation} 
   \begin{array}{r}
   \texttt{fvixx0(cmplx* fi, cmplx* vc, float* gal,} \\
   \texttt{cmplx* fvi)} 
    \end{array}
  \end{equation}
  where the inputs and the outputs are:
  \begin{equation}
   \begin{array}{ll}
    \textsc{Inputs:} & \\
     \texttt{cmplx fi[6]} & \textrm{flowing-In fermion wavefunction} \\
     \texttt{cmplx vc[6]} & \textrm{vector wavefunction} \\
     \texttt{float gal[2]} & \textrm{coupling constants of the
      \texttt{FFV} vertex}  \\ \\
    \textsc{Outputs:} & \\
     \texttt{cmplx fvi[6]} & \textrm{off-shell fermion wavefunction} \\
                         & \texttt{|f',vc,fi>} 
    \end{array}
  \end{equation}

  \subsubsection{\texttt{fvoxx0}}
  \label{app:heget-ffv-fvo}

  The function \texttt{fvoxx0} (List.~\ref{list:fvoxx0})
  computes the off-shell massless fermion wavefunction
  from a flowing-Out external fermi\-on and a vector boson.
  It has the arguments:
  \begin{equation} 
   \begin{array}{r}
    \texttt{fvoxx0(cmplx* fo, cmplx* vc, float* gal,} \\
    \texttt{cmplx* fvo)} 
   \end{array}
  \end{equation}
  where the inputs and the outputs are:
  \begin{equation}
   \begin{array}{ll}
    \textsc{Inputs:} & \\
     \texttt{cmplx fo[6]} & \textrm{flowing-Out fermion wavefunction} \\
     \texttt{cmplx vc[6]} & \textrm{vector wavefunction} \\
     \texttt{float gal[2]} & \textrm{coupling constants of the
      \texttt{FFV} vertex}  \\ \\
    \textsc{Outputs:} & \\
     \texttt{cmplx fvo[6]} & \textrm{off-shell fermion wavefunction} \\
                         & \texttt{<fo,vc,f'|} 
    \end{array}
  \end{equation}

  \subsubsection{\texttt{jioxx0}}
  \label{app:heget-ffv-jio}

  This function, \texttt{jioxx0} (List~\ref{list:jioxx0})
  computes the off-shell vector wavefunction from an external
  fermion pair. 
  The off-shell vector boson wavefunction is given in the Feynman gauge
  for massless vectors (photons and gluons in the SM).
  It has the arguments:
  \begin{equation} 
   \begin{array}{r}
   \texttt{jioxx0(cmplx* fi, cmplx* fo, float* gal,} \\
   \texttt{cmplx* jio)} 
   \end{array}
  \end{equation}
  where the inputs and the outputs are:
  \begin{equation}
   \begin{array}{ll}
    \textsc{Inputs:} & \\
     \texttt{cmplx fi[6]} & \textrm{flowing-In fermion wavefunction} \\
     \texttt{cmplx fo[6]} & \textrm{flowing-Out fermion wavefunction} \\
     \texttt{float gal[2]} & \textrm{coupling constants of the
      \texttt{FFV} vertex}  \\ \\
    \textsc{Outputs:} & \\
     \texttt{cmplx jio[6]} & \textrm{vector current $j^{\mu}$
      (\texttt{<fo|V|fi>})} 
    \end{array}
  \end{equation}

\begin{lstlisting}[caption=iovxxx.cu,label=list:iovxxx]
#include "cmplx.h"

__device__
void iovxx0(cmplx* fi, cmplx* fo, cmplx* vc,
            float* gal, cmplx& vertex)
{
    vertex = 
        gal[0]*((fo[2]*fi[0]+fo[3]*fi[1])*vc[0]
                +(fo[2]*fi[1]+fo[3]*fi[0])*vc[1]
                -((fo[2]*fi[1]-fo[3]*fi[0])*vc[2])
                *mkcmplx(0.0f, 1.0f)
                +(fo[2]*fi[0]-fo[3]*fi[1])*vc[3])
        +gal[1]*((fo[0]*fi[2]+fo[1]*fi[3])*vc[0]
                 -(fo[0]*fi[3]+fo[1]*fi[2])*vc[1]
                 +((fo[0]*fi[3]-fo[1]*fi[2])*vc[2])
                 *mkcmplx(0.0f, 1.0f)
                 -(fo[0]*fi[2]-fo[1]*fi[3])*vc[3]);
    return;
}
\end{lstlisting}
 
\begin{lstlisting}[caption=fvixx0.cu,label=list:fvixx0]
#include "cmplx.h"

__device__
void fvixx0(cmplx* fi, cmplx* vc, float* gal,
            cmplx* fvi)
{
    fvi[4] = fi[4]-vc[4];
    fvi[5] = fi[5]-vc[5];

    float  pf[4];
    pf[0] = fvi[4].re;
    pf[1] = fvi[5].re;
    pf[2] = fvi[5].im;
    pf[3] = fvi[4].im;

    float pf2 = pf[0]*pf[0] - 
        (pf[1]*pf[1] + pf[2]*pf[2] + pf[3]*pf[3]);
    cmplx  cI = mkcmplx(    0.0f,    1.0f);
    float   d = -1.0f/pf2;

    cmplx sl1 =  (vc[0] + vc[3])*fi[0] 
        + (vc[1] - cI*vc[2])*fi[1];
    cmplx sl2 =  (vc[0] - vc[3])*fi[1] 
        + (vc[1] + cI*vc[2])*fi[0];
    cmplx sr1 =  (vc[0] - vc[3])*fi[2] 
        - (vc[1] - cI*vc[2])*fi[3];
    cmplx sr2 =  (vc[0] + vc[3])*fi[3] 
        - (vc[1] + cI*vc[2])*fi[2];

    fvi[0] = ( gal[0]*((pf[0]-pf[3])*sl1 
                       -  conj(fvi[5])*sl2))*d;
    fvi[1] = ( gal[0]*((pf[0]+pf[3])*sl2 
                       -       fvi[5] *sl1))*d;
    fvi[2] = ( gal[1]*((pf[0]+pf[3])*sr1 
                       +  conj(fvi[5])*sr2))*d;
    fvi[3] = ( gal[1]*((pf[0]-pf[3])*sr2 
                       +       fvi[5] *sr1))*d;
    return;
}
\end{lstlisting}
\begin{lstlisting}[caption=fvoxx0.cu,label=list:fvoxx0]
#include "cmplx.h"

__device__
void fvoxx0(cmplx* fo, cmplx* vc, float* gal,
            cmplx* fvo)
{
    fvo[4] = fo[4]+vc[4];
    fvo[5] = fo[5]+vc[5];

    float  pf[4];
    pf[0] = fvo[4].re;
    pf[1] = fvo[5].re;
    pf[2] = fvo[5].im;
    pf[3] = fvo[4].im;

    float pf2 = pf[0]*pf[0] 
        - (pf[1]*pf[1] + pf[2]*pf[2] + pf[3]*pf[3]);
    cmplx  cI = mkcmplx(    0.0f,    1.0f);
    float   d = -1.0f/pf2;
    
    cmplx sl1 =  (vc[0] + vc[3])*fo[2] 
        + (vc[1] + cI*vc[2])*fo[3];
    cmplx sl2 =  (vc[0] - vc[3])*fo[3] 
        + (vc[1] - cI*vc[2])*fo[2];
    cmplx sr1 =  (vc[0] - vc[3])*fo[0] 
        - (vc[1] + cI*vc[2])*fo[1];
    cmplx sr2 =  (vc[0] + vc[3])*fo[1] 
        - (vc[1] - cI*vc[2])*fo[0];

    fvo[0] = ( gal[1]*((pf[0]+pf[3])*sr1 
                       +      fvo[5] *sr2 ))*d;
    fvo[1] = ( gal[1]*((pf[0]-pf[3])*sr2 
                       + conj(fvo[5])*sr1 ))*d;
    fvo[2] = ( gal[0]*((pf[0]-pf[3])*sl1 
                       -      fvo[5] *sl2 ))*d;
    fvo[3] = ( gal[0]*((pf[0]+pf[3])*sl2 
                       - conj(fvo[5])*sl1 ))*d;

    return;
}
\end{lstlisting}

\begin{lstlisting}[caption=jioxx0.cu,label=list:jioxx0]
#include "cmplx.h"

__device__
void jioxx0(cmplx* fi, cmplx* fo, float* gal,
            cmplx* jio)
{
    jio[4]=fo[4]-fi[4];
    jio[5]=fo[5]-fi[5];

    float DD=1.0f/(
          (jio[4].re)*(jio[4].re) - (jio[5].re)*(jio[5].re) 
        - (jio[5].im)*(jio[5].im) - (jio[4].im)*(jio[4].im));

    jio[0]=(gal[0]*(fo[2]*fi[0]+fo[3]*fi[1])
            +gal[1]*(fo[0]*fi[2]+fo[1]*fi[3]))*DD;

    jio[1]=(gal[1]*(fo[0]*fi[3]+fo[1]*fi[2])
            -gal[0]*(fo[2]*fi[1]+fo[3]*fi[0]))*DD;

    jio[2]=(gal[0]*(fo[2]*fi[1]-fo[3]*fi[0])
            -gal[1]*(fo[0]*fi[3]-fo[1]*fi[2]))
        *mkcmplx(0.0f,DD);

    jio[3]=(gal[1]*(fo[0]*fi[2]-fo[1]*fi[3])
            -gal[0]*(fo[2]*fi[0]-fo[3]*fi[1]))*DD;

    return;
}
\end{lstlisting}

 \subsection{Complex numbers}
 \label{app:heget-complex}
\begin{lstlisting}[caption=cmplx.h,label=list:cmplx]
#ifndef __cmplx_h__
#define __cmplx_h__

typedef struct __align__(8){
   float re;
   float im;
} cmplx;

inline __host__ __device__ 
cmplx mkcmplx(float re, float im){
    cmplx z; 
    z.re=re; 
    z.im=im; 
    return z;
}

inline __host__ __device__ 
cmplx mkcmplx(cmplx z){
    return mkcmplx(z.re,z.im);
}

inline __host__ __device__ 
cmplx mkcmplx(float a){
    return mkcmplx(a,0.0);
}

inline __host__ __device__  
float real(cmplx a){
    return a.re;
}

inline __host__ __device__  
float imag(cmplx a){
    return a.im;
}

inline __host__ __device__ 
cmplx conj(cmplx a){
    return mkcmplx(a.re,-a.im);
}

inline __host__ __device__  
cmplx operator+(cmplx a, cmplx b){
    return mkcmplx(a.re + b.re, a.im + b.im);
}

inline __host__ __device__  
void operator+=(cmplx &a, cmplx b){
    a.re += b.re; a.im += b.im;
}

inline __host__ __device__  
cmplx operator+(cmplx a){
    return mkcmplx(+a.re, +a.im);
}

inline __host__ __device__  
cmplx operator-(cmplx a, cmplx b){
    return mkcmplx(a.re - b.re, a.im - b.im);
}

inline __host__ __device__  
void operator-=(cmplx &a, cmplx b){
    a.re -= b.re; a.im -= b.im;
}

inline __host__ __device__  
cmplx operator-(cmplx a){
    return mkcmplx(-a.re, -a.im);
}

inline __host__ __device__  
cmplx operator*(cmplx a, cmplx b){
    return mkcmplx((a.re * b.re) - (a.im * b.im),
                   (a.re * b.im) + (a.im * b.re));
}

inline __host__ __device__  
cmplx operator*(cmplx a, float s){
    return mkcmplx(a.re * s, a.im * s);
}

inline __host__ __device__  
cmplx operator*(float s, cmplx a){
    return mkcmplx(a.re * s, a.im * s);
}

inline __host__ __device__  
void operator*=(cmplx &a, float s){
    a.re *= s; a.im *= s;
}

inline __host__ __device__  
cmplx operator/(cmplx a, cmplx b){
    float tmpD=(1.0f/(b.re*b.re+b.im*b.im));
    return mkcmplx( 
        ( (a.re * b.re) + (a.im * b.im))*tmpD,
        (-(a.re * b.im) + (a.im * b.re))*tmpD
        );
}

inline __host__ __device__  
cmplx operator/(cmplx a, float s){
    float inv = 1.0f / s;
    return a * inv;
}

inline __host__ __device__  
cmplx operator/(float s, cmplx a){
    float inv = s*(1.0f/(a.re*a.re+a.im*a.im));
    return mkcmplx(inv*a.re,-inv*a.im);
}

inline __host__ __device__  
void operator/=(cmplx &a, float s){
    float inv = 1.0f / s;
    a *= inv;
}

inline __host__ __device__ 
float fabsc(cmplx a){
    return sqrtf((a.re*a.re)+(a.im*a.im));
}

inline __host__ __device__ 
float fabs2c(cmplx a){
    return (a.re*a.re)+(a.im*a.im);
}
#endif
\end{lstlisting}

 \section{A sample code for unrolling the permutation loop for three
 photon production process}\label{app:sample-unroll}

\begin{lstlisting}[caption=uux3a(unrolled).cu,label=list:uux3aunroll]
        cmplx w[5][6];

        ixxxx1(p1, nh1, +1, w[0]);
        oxxxx2(p2, nh2, -1, w[1]);
        vxxxx0(p3, nh3, +1, w[2]);
        vxxxx0(p4, nh4, +1, w[3]);
        vxxxx0(p5, nh5, +1, w[4]);

        cmplx w01[6],w02[6];
        
        cmplx ampsum = mkcmplx(0.0f, 0.0f);
        
        int Flag_Perm=1;
        int a[3];
        a[0] = 2;
        a[1] = 3;
        a[2] = 4;
        
        do {
            cmplx amp;
            fvoxx0(w[1], w[a[0]], gau, w09);
            fvoxx0(w09 , w[a[1]], gau, w10);
            iovxx0(w[0], w10, w[a[2]], gau, amp);
            ampsum += amp;
            fvoxx0(w[1], w[a[0]], gau, w09);
            fvoxx0(w09 , w[a[2]], gau, w10);
            iovxx0(w[0], w10, w[a[1]], gau, amp);
            ampsum += amp;
            fvoxx0(w[1], w[a[2]], gau, w09);
            fvoxx0(w09 , w[a[0]], gau, w10);
            iovxx0(w[0], w10, w[a[1]], gau, amp);
            ampsum += amp;
            iPNext(a, 2, Flag_Perm);
        } while (Flag_Perm>0);
\end{lstlisting}

  %
  %

\end{document}